# Guidelines for Correlative Imaging and Analysis of Reactive Lithium Metal Battery Materials

Shuang Bai[1, 2, +], Zhao Liu[3, +], Diyi Cheng[1], Bingyu Lu[4], Nestor J. Zaluzec[2], Ganesh Raghavendran[4], Shen Wang[4], Thomas S. Marchese[2], Brandon van Leer[3], Letian Li[3], Lin Jiang[3], Adam Stokes[3], Joseph P. Cline[3], Rachel Osmundsen[3], Paul Barends[3], Alexander Bright[3],
Minghao Zhang[2, *], Ying Shirley Meng[2, 4, 5, *]

1. Program of Materials Science and Engineering, University of California San Diego, La Jolla, California 92093, United States.
2. Pritzker School of Molecular Engineering, University of Chicago, Chicago, Illinois 60637, United States.
3. Thermo Fisher Scientific, 5350 NE Dawson Creek Drive, Hillsboro, Oregon 97124, United States.
4. Aiiso Yufeng Li Family Department of Chemical and Nano Engineering, University of California San Diego, La Jolla, California 92093, United States.
5. Argonne National Laboratory, Lemont, Illinois 60439, United States.

[+]These authors contributed equally to the work.
Corresponding authors: miz016@uchicago.edu, shirleymeng@uchicago.edu

# Summary

To unlock the full potential of lithium metal batteries, a deep understanding of lithium metal's reactivity and its solid electrolyte interphase (SEI) is essential. Correlative imaging, combining focused ion beam (FIB) and electron microscopy (EM), offers a powerful approach for multi-scale characterization. However, the extreme reactivity of lithium metal and its SEI presents challenges in investigating deposition and stripping mechanisms. In this work, we systematically evaluated the storage stability of lithium metal in a glovebox (Ar atmosphere, <0.1 ppm moisture and oxygen) before and after electrochemical deposition. We then assessed different FIB ion sources ($Ga^+$, $Xe^+$, $Ar^+$) for their impact on lithium metal lamella preparation for transmission electron microscopy (TEM). Furthermore, we examined cryogenic-TEM (cryo-TEM) transfer methods, optimizing for minimal contamination during sample handling. Contrary to prior assumptions, we demonstrate that high resolution imaging of pure lithium metal at room temperature is achievable using inert gas sample transfer (IGST) with an electron dose rate exceeding $10^3$ e/Å²·s, without significant detectable damage. In contrast, SEI components, such as $Li_2CO_3$, and LiF, display much greater sensitivity to electron beams, requiring cryogenic conditions and precise dose control for nano/atomic scale imaging. We quantified electron dose limits for these SEI compounds to track their structural evolution under irradiation. Based on these findings, we propose a robust protocol for lithium metal sample handling—from storage to atomic-level characterization—minimizing damage and contamination. This work paves the way for more accurate and reproducible studies, accelerating the development of next-generation lithium metal batteries by ensuring the preservation of native material properties during analysis.

## Introduction

Lithium metal batteries have emerged as a promising energy storage technology with the potential to revolutionize portable electronics and electric vehicles.[1] With its significantly higher specific capacity of 3860 mAh/g and lower electrochemical potential (-3.04 V vs. the standard hydrogen electrode), Li metal presents an opportunity for achieving higher energy density batteries compared to those utilizing graphite anodes.[2] This is particularly important for applications where maximizing energy storage capacity in a limited space is critical, such as in electric vehicles, drones and portable electronics. Despite these advantages, there are challenges associated with the use of lithium metal anodes, including short cycle life and safety issues caused by dendrite growth and inactive lithium formation.

The morphology and composition of electrochemically deposited lithium are widely recognized as a critical factor influencing the cycle life and safety of lithium metal batteries.[3] Additionally, due to the high reduction potential of lithium, the solid electrolyte interphase (SEI) formed between the lithium metal and electrolyte functions as a passivation layer. This SEI layer serves to protect the lithium metal surface by mitigating continuous side reactions with the electrolyte while also ensuring good Li-ion conductivity. Consequently, the properties of the SEI play a crucial role in governing the lithium deposition and stripping processes,[4,5] positioning lithium metal and the SEI as the key components that most critically impact the performance of lithium metal batteries. Consequently, the properties of the SEI significantly influence the deposition and stripping processes of lithium,[4,5] making lithium metal and SEI the most critical components affecting the performance of lithium metal batteries. Comprehensive characterization techniques and advanced analytical methods are thus essential for elucidating the intricate nature of the SEI and its relationship with lithium metal and thus inform the development of innovative strategies aimed at controlling the deposition and stripping processes of lithium metal.[6]

The electrochemical lithium metal plating and stripping involve coupled dynamics include electron transport, ion diffusion, interphase formation, interfacial evolution, and structural transformations across multiple length scales.[7–9] Correlative tools to obtain structural and chemical information across micro- and nano-scales are essential for understanding these dynamics.[10] Due to the high reactivity of both lithium metal and the SEI with environmental factors and external probes, maintaining sample integrity during correlative characterization across multiple analytical tools and length scales poses a significant challenge. Furthermore, the battery field currently lacks established metrological guidelines for thoroughly investigating these reactive materials across various length scales while preserving their native states. This challenge remained unresolved until 2017, when the introduction of cryo-electron microscopy (Cryo-EM) offered a significant breakthrough. By stabilizing lithium metal samples below -170°C in a high vacuum environment ($< 10^{-4}$ Pa), contamination from moisture and oxygen can be minimized, and electron beam-induced damage to the lithium samples is greatly mitigated. Under these controlled conditions, high-resolution imaging of lithium metal became achievable for the first time. Notably, cryogenic methods hold the potential to stabilize weakly bonded materials and reactive interfaces which typically degrade under high-energy electron beam irradiation and environmental exposure. These advanced cryogenic imaging methods have been recently applied to study non-biological irradiation sensitive materials such as the lithium metal anode and the SEI in batteries[10–14] and the structure of lithium metal anodes and their nanostructured SEI species has been resolved using cryogenic transmission electron microscopy (cryo-TEM).[11,12,15] Furthermore, cryogenic-focused ion beam techniques (cryo-FIB) have been introduced to the battery field, enabling the examination of electrodeposited lithium metal at the scale of hundreds of microns.[14] Importantly, cryo-FIB allows for the precise thinning of lithium metal samples to below 100 nm while minimizing material damage, thereby making them suitable for detailed TEM analysis.

The integration of cryo-TEM and cryo-FIB technologies for correlative imaging has revitalized efforts towards developing practical metal anode batteries with improved safety and prolonged cyclability.[16,17] However, when, where, and how to employ cryogenic imaging techniques in battery studies remain topics of debate and there is need for a proper characterization protocol. Recent reports have demonstrated atomic resolution TEM imaging of lithium metal without inducing irradiation damage, even at room temperature, with a dose rate exceeding 1000 e/Å$^2$·s.[18] This finding challenges the previously held belief that cryo-EM is indispensable for lithium metal imaging.[11,12] Additionally, the plasma focused ion beam (PFIB) specimen preparation technique using Xe$^+$/ Ar$^+$ ions has been reported to avoid introducing discernible morphological changes or ion-induced contamination which is found in traditional focused ion beam instruments employing Ga$^+$.[19] The complex nature of cryo-TEM characterization also poses a significant challenge, as it allows for the generation of discrepancies at various stages of operation due to diverse sample handling protocols which are being used by a significant portion of the community without standardization. These uncertainties are particularly prevalent during the entire sample handling and transfer procedures.

This study introduces a correlative imaging approach tailored for reactive battery materials, using lithium metal and its SEI components as case studies. First, we monitored the lithium metal stored in the Ar glovebox, which had been prepared using various methods, and determined the optimal storage time window for conducting further experiments. Next, we assessed the effects of different focused ion beam (FIB) sources on the preparation of lithium specimens for TEM imaging. Additionally, we evaluated the impact of different cryo-EM transfer systems on sample integrity. Finally, we examined the influence of imaging dose and cryogenic conditions on the stability of the lithium samples during imaging. Based on this compendium of knowledge, we propose, herein, a robust protocol for lithium metal study from storage to atomic-level characterization. Our findings suggest a reconsideration of two prevailing assumptions in cryo-EM research on battery materials: firstly, that beam dose control is unnecessary as long as imaging is conducted at cryogenic temperatures; and secondly, the prevalent belief that $Li_2O$ is the dominant SEI component irrespective of the electrolyte applied. We show that both of these assumptions are mistaken. This work advances the understanding of lithium metal and SEI reactivity, providing new insights into their behavior. It also establishes a robust framework for more accurate and reliable characterization of reactive battery materials, paving the way for future innovations in lithium battery technology.

## Results & Discussion

**Storage and FIB Preparation of Lithium Metal specimens for EM Studies**

A critical issue we needed to address at the beginning of this study was how to properly store metal electrodes after they have undergone cycling in a battery. In the field of battery research, the standard approach is to disassemble the lithium metal anode from the battery in an inert atmosphere glovebox and then store it in the same controlled environment. However, a key question arises: does this stored anode have a 'shelf life'? Even in an Ar-filled glovebox (<0.1 ppm oxygen and moisture), residual traces of water and oxygen exist and will react with the lithium metal electrode over time. As a result, the progressive formation of $Li_2O$, LiOH, and $Li_2CO_3$ can compromise the chemical integrity of the stored sample. More critically, after cycling in the battery, changes in surface morphology—such as increased roughness, surface area, and composition—may make the lithium metal even more susceptible to these reactions. Therefore, it is likely that lithium metal electrodes, prepared under different conditions, indeed have a limited shelf life. To validate this shelf life in a quantitively perspective, we employed titration gas chromatography (TGC) to quantify the active lithium metal inventory change upon storage inside an Ar-filled glovebox. This method was used to quantitatively analyze the $Li^0$ content in various lithium samples.[1] Three types of lithium samples were analyzed: commercial lithium foil (Commercial Lithium), electrochemically deposited lithium with Bi-salt electrolyte (4.7M LiFSI and 2.3M LiTFSI in DME), and electrochemically deposited lithium with Gen 2 electrolyte (1.2M $LiPF_6$ in EC: EMC). These samples were quantified by TGC after different storage durations inside an Ar-filled glovebox. Interestingly, as shown in **Figure 1A**, lithium samples prepared by different methods exhibit significantly different environmental reactivity within the glovebox, despite its Ar atmosphere containing <0.1 ppm oxygen and moisture. After 7 days, the commercial lithium showed less than 10% lithium inventory loss, while the Gen 2 sample experienced the most substantial loss, with over 40% of the lithium inventory depleted. The Bi-Salt sample fell between these two extremes, with less than 20% of its lithium inventory lost. These differences can be attributed to variations in microstructure as well as the SEI composition, which greatly influence the environmental stability of the lithium samples.[2,20] The commercial lithium sample, with its dense, compact surface passivation layer and uniform bulk, demonstrated greater resistance to residual contaminants. In contrast, electrochemically deposited lithium, with its dendritic formations, organic-rich SEI, and non-uniform deposition, exhibited significantly higher reactivity, leading to substantial lithium inventory loss. Electrolytes with better cycling performance, like the bi-salt electrolyte, can mitigate these non-uniform depositions and organic-rich SEI, resulting in better glovebox storage stability compared to conventional electrolytes such as Gen 2.

Based on these results, we recommend that after disassembling the batteries for analysis, any FIB Lithium specimen preparation processes should ideally begin within 1-2 days. This minimizes the loss of lithium metal due to glovebox contamination to less than 10%. For lithium metal materials cycled with electrolytes exhibiting low coulombic efficiency, this time window should be reduced to just a few hours after disassembly. Ideally, for laboratories equipped with TGC, conducting similar assessments as outlined in our study will allow them to determine the shelf life of their specific electrode materials within their own glovebox environments.

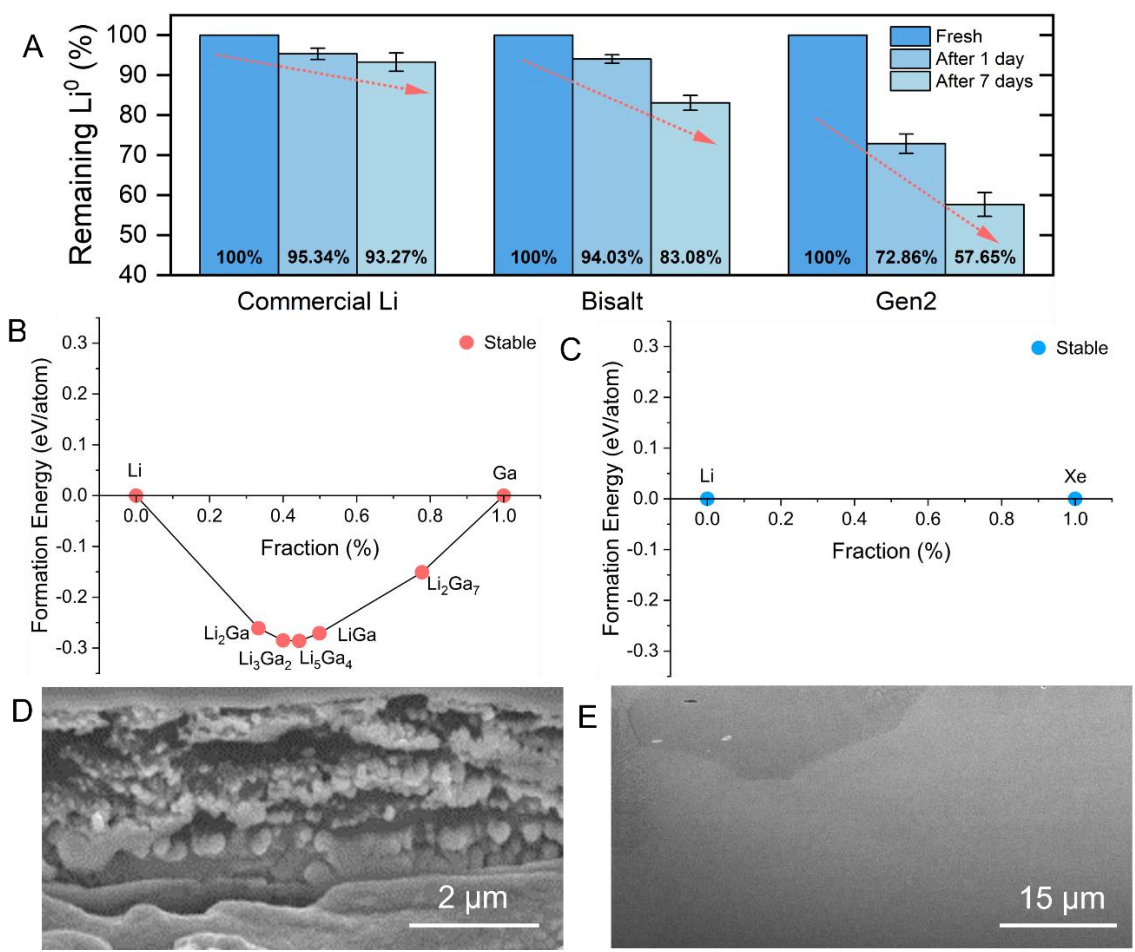

**Figure 1. Storage test of different lithium metals in an Ar-filled glovebox and evaluation of the ion beam induced morphological and compositional change of Lithium Metal in Ga$^+$ FIB and Xe$^+$ PFIB.** (**A**) Comparison of Li metal inventory loss, tested by titration gas chromatography (TGC), among commercial lithium, electrochemically deposited lithium in Bisalt electrolyte (0.25 mAh/cm$^2$ at 2 mA/cm$^2$, 4.7M LiFSI and 2.3M LiTFSI in DME), and electrochemically deposited lithium in Gen2 electrolyte (0.25 mAh/cm$^2$ at 2 mA/cm$^2$, 1.2M LiPF$_6$ in EC:EMC). Calculated alloy phases of (**B**) Li-Ga system, and (**C**) Li-Xe system. The phase diagram is constructed based on the data from the Materials Project. Cross-section images of commercial lithium foil ion milled at room temperature using (**D**) Ga$^+$ FIB with a 3nA milling current, and (**E**) Xe$^+$ PFIB with a 1uA milling current.

With the lithium anode within the proper stored time window, next, before performing TEM experiments and obtaining high-quality images, the preparation of TEM specimens (lamella) by FIB is crucial. Here, several considerations must be addressed: (1) The sample thickness should be as close to or less than 100 nm for comprehensive microstrucutral analysis in the TEM. (2) The preparation process should minimize any damage to the sample's morphology and composition. (3) Environmental contamination during the preparation process should be avoided as much as possible. Given these factors, using FIB is generally one of the few reasonable choices for Li TEM sample preparation. FIB allows for high-resolution processing of lithium metal samples, ensuring that the final sample thickness is sufficiently thin. Additionally, since the preparation can be performed in a high-vacuum environment, it can, with due care, minimize exposure to gases and moisture, thereby reducing the risk of environmental contamination.

In FIB, low angle ion milling involves substantial ion beam collisions with the sample, leading to the transfer of kinetic energy and the resulting specimen thinning (i.e. spallation of the specimen) from impact events on its surface. When conventional $Ga^+$ ion FIB milling is performed on commercial lithium foil at room temperature, significant morphological damage occurs as illustrated in **Figure 1D**, at beam currents as low as 3 nA at 30 k**V**. Using a significantly lower ion current, such as several pA, could potentially mitigate beam-induced damage. Unfortunately, this approach becomes impractical due to the prohibitive time required to ion beam thin the specimen under such low current conditions, as this extends the preparation time to dozens of hours thus increasing the chance for other deleterious events. This finding aligns with our prior observations, highlighting the necessity of cryogenic conditions and PFIB operations to mitigate adverse side effects with acceptable experimental time window.[14] Various mechanisms have been proposed to elucidate the mitigation effects observed during cryogenic sample milling. Firstly, at room temperature and under relatively high vacuum (~$10^{-4}$ Pa) in the FIB chamber, bulk lithium metal is susceptible to local Joule heating, leading to melting and evaporation, which cryogenic operations can minimize. Secondly, in the $Ga^+$ FIB substantial ion implantation occurs to depths >500 nm when the FIB is operated at 30kV. This allows Ga-rich compounds to form, potentially causing sample deformation as well as Li-Ga phase formation. The use of an inert gas plasma FIB (PFIB), utilizing either Xe or Ar plasma as the ion source for milling materials, reduces this due to the lower reactivity of both Xe and Ar with lithium metal. This is elucidated in **Figure 1B**, where our calculations indicate that a series of Ga-Li alloys can form over a wide range of elemental ratios, in contrast we see that Xe remains inert with lithium metal (**Figure 1C**). To further evaluate the reactivity of $Ga^+$ and $Xe^+$ ion beams with lithium metal at room temperature, both ion sources were experimentally used to ion-mill pristine lithium metal. The milling current in the $Xe^+$ PFIB was two orders of magnitude higher than the equivalent experiment done in the $Ga^+$ FIB, which would have subjected the lithium metal sample to considerable Joule heating. However, the morphology of the lithium metal in these experiments remained unchanged, suggesting that, at room temperature, lithium metal morphology is not sensitive to Joule heating. This indicates that the primary factor influencing lithium metal morphological changes appears to be the reactivity between the ion source material and the lithium metal itself. In addition, the retention of distinct grain boundaries observed after $Xe^+$ PFIB milling between lithium grains manifests the ability of $Xe^+$ PFIB to preserve crystalline specimen characteristics for scanning electron microscopy (SEM) as well as TEM examinations. The sharp morphology differences illustrated in **Figure 1D** and **1E** highlight the critical role of chemical reactivity between lithium metal and $Ga^+$ ions as the primary cause of bulk structural damage, a phenomenon as we have shown is mitigated under cryogenic PFIB conditions. The comparable effect of inert $Ar^+$ PFIB milling in maintaining lithium metal cross-section morphology at room temperature provides further validation of the chemical reactivity driven reaction mechanism (**Figure S1**).

In addition to evaluating the ion beam-induced morphological effects, the compositional changes in lithium metal resulting from (P)FIB milling were also characterized. X-ray Energy Dispersive Spectroscopy (XEDS) integrated into the (P)FIB instruments, which were employed for these studies, was used to monitor changes in the nominal composition of lithium specimen. As alluded to previously, the higher reactivity between Ga and Li leads to an implantation of Ga in the specimen which is readily detectable using XEDS (**Figure S2**), while similar experiments conducted in the $Xe^+$ PFIB demonstrated nearly undetectable Xe signal.

Regardless of the ion source, surface contamination from oxygen is unavoidable at room temperature, even when experiments are conducted under oxygen partial pressures below $10^{-5}$ Pa. A recorded video (**Video S1**) captures the sequence of contamination on a pristine lithium metal surface upon exposure within the PFIB chamber. This surface contamination can be moderated by performing PFIB milling under cryogenic conditions, as demonstrated in **Figure S3**. These findings indicate that while PFIB operation at room temperature with an inert ion source is sufficient for bulk specimen preparation and characterization of reactive metals, cryogenic milling will be essential for accurately investigating interfacial chemistry.

The environmental sensitivity of lithium metal poses significant challenges for both storage and specimen preparation. Fortunately, by carefully selecting the storage time window, ion source, and TEM lamella preparation temperature, the lithium samples' integrity can largely be preserved.

**Lamella Sample Transfer for TEM Imaging**

Next, we will focus on discussing the significance of sample transfer, which consists of two key stages: (1) transferring the lithium sample from the glovebox to the (P)FIB, and subsequently back into the glovebox after lamella preparation; and (2) transferring the prepared lamella from the glovebox to the TEM. In the first stage, our previous work demonstrated that the CleanConnect Shutter system effectively minimizes exposure to air and moisture. This device, developed by ThermoFisher Scientific, transfers the sample to the PFIB-SEM for lamella preparation, with overpressure Ar gas providing protection against air exposure.

For the second transfer stage, which involves transferring the FIB prepared lithium lamella to the TEM system, selecting an appropriate cryo-TEM holder is crucial. So far, two types of commercially available holder types enable cryo-TEM characterization: the cryo-transfer holder and the cooling holder.[21] The cryo-transfer holder is designed to pre-cool the holder before loading the cryogenically prepared sample grid onto it while emersed in liquid nitrogen.[22,23] This method is widely employed for biological samples, as it effectively maintains the sample integrity in a frozen cryogenic state and reduces sample reactivity. However, during the process of loading the grid onto the holder and transferring the sample to TEM, some frozen water inevitably forms on the sample (**Figure 2A**). Therefore, controlling humidity in the area where the TEM and holder are located is necessary to alleviate the icing issue associated with the cryo-transfer method. This requirement further limits the available instrumentation options, making it more challenging to find suitable equipment for these sensitive procedures. We also note that a secondary (smaller source) of water/ice during cryo-EM studies is condensation from the vacuum system of the instrument and thus the use of "cryo-fingers" in the microscope system should always be employed during any cryogenic operations. On the other hand, the cooling holder (**Figure 2B**) also allows for the cooling of the sample after inserting the holder into the TEM column. This method requires protecting the sample in an inert gas environment during all steps prior to observation in high vacuum and is typically accomplished using a glovebag system surrounding the holder with an inert gaseous environment to enable transfer between the glove bag and the TEM. However, to facilitate insertion into the vacuum system of any instrument, a few seconds exposure of the sample to air is unavoidable because the cooling holder is not fully isolated from the ambient environment. Any short exposure of samples to air could lead to changes in sample surface morphology and chemistry due to contamination. An example of which is shown in **Figure S4**. Here, with only 15 seconds of exposure to air, the scanning transmission electron microscopy (STEM) dark field and high angular dark field images show significant structural changes of TEM specimen. The particulate and network-shaped contrast on the surface of the lamella is indicative of oxidation or carbonation (**Figure 2B**).

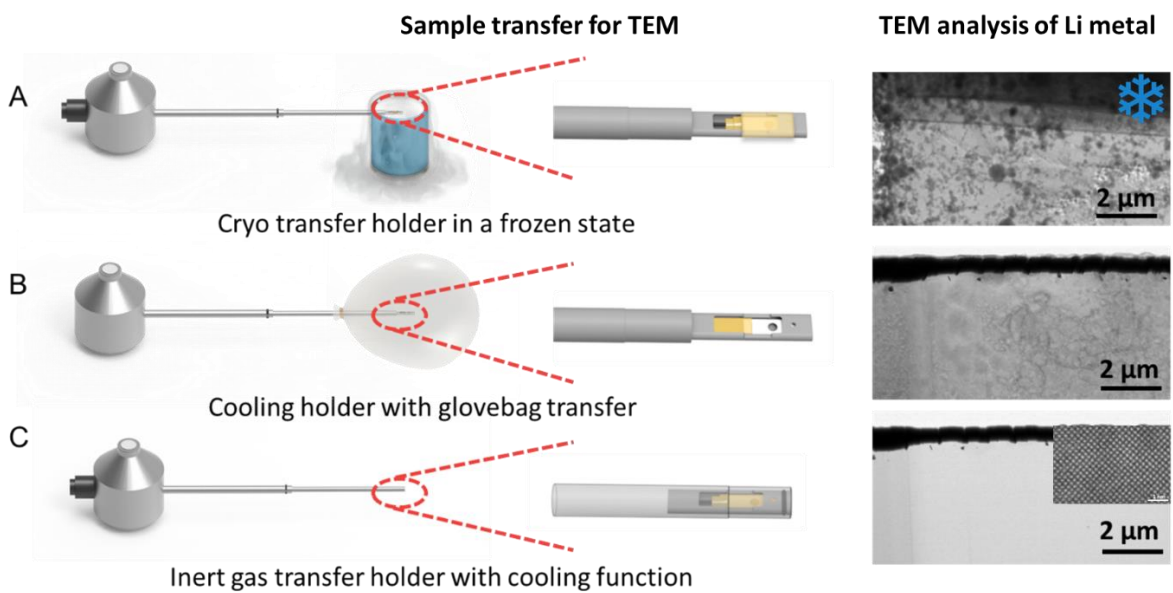

**Figure 2.** Correlative imaging workflow for TEM through different sample transfer methods to observe reactive lithium metal down to the atomic scale: (**A**) Cryo transfer holder, (**B**) Cooling holder with glovebag transfer, (**C**) Inert gas sample transfer holder with cooling function.

These two examples suggest that an ideal sample transfer for Li characterization in the TEM should be as complete as possible to avoid air exposure or ice formation. To realize these requirements, we employed an inert gas sample transfer holder (IGST). As illustrated in **Figure 2C**, this design of holder generally incorporates a retractable tip that effectively seals the sample with an O-ring and since this can be accomplished within the inert atmosphere glovebox, the seal largely safeguards the specimen from exposure to ambient air prior to observation in the TEM. Ice formation is also minimized as the sample is transferred at room temperature from a low moisture environment. After inserting the IGST holder into the TEM vacuum system, the sealed holder can be safely opened to perform further experiments. As **Figure 2C** illustrates, under these conditions there is minimal surface contamination of the lithium specimens transferred when employing an IGST holder. Moreover, the direct comparison results of lithium metal lamella transfer via Ar gas protection versus transfer with 15 seconds of air exposure are illustrated in **Figure S4** and **S5**, which demonstrate the IGST holder transfer maintains the integrity of the lithium and high resolution TEM images of lithium metal can be obtained at room temperature (**Figure 2C**). The result demonstrates the imaging of lithium metal at room temperature. More importantly, even though a high dose rate of $3.5 \times 10^3$ e/Å$^2$·s is applied, there is no discernible irradiation damage. This outcome validates the effectiveness of the protocol for correlative imaging of lithium metal with minimal artifacts across multiple instruments.

**Protocols for Mitigating the Reactivity of SEI to Electron Beam**

In previous studies, cryo-TEM has identified several inorganic components in the SEI of electrochemically deposited lithium, including LiF, $Li_2CO_3$, and $Li_2O$.[24–27] Although these studies have recognized the electron-sensitive nature of SEI components, detailed imaging conditions, including electron beam dose limitations, have rarely been reported.[15] In this study we also quantify electron irradiation damage process of these SEI components at different temperatures. This is a critical step for studying the mechanism of irradiation damage in electron-reactive battery material specimens because it determines what dosage parameters are available for minimizing the damage.

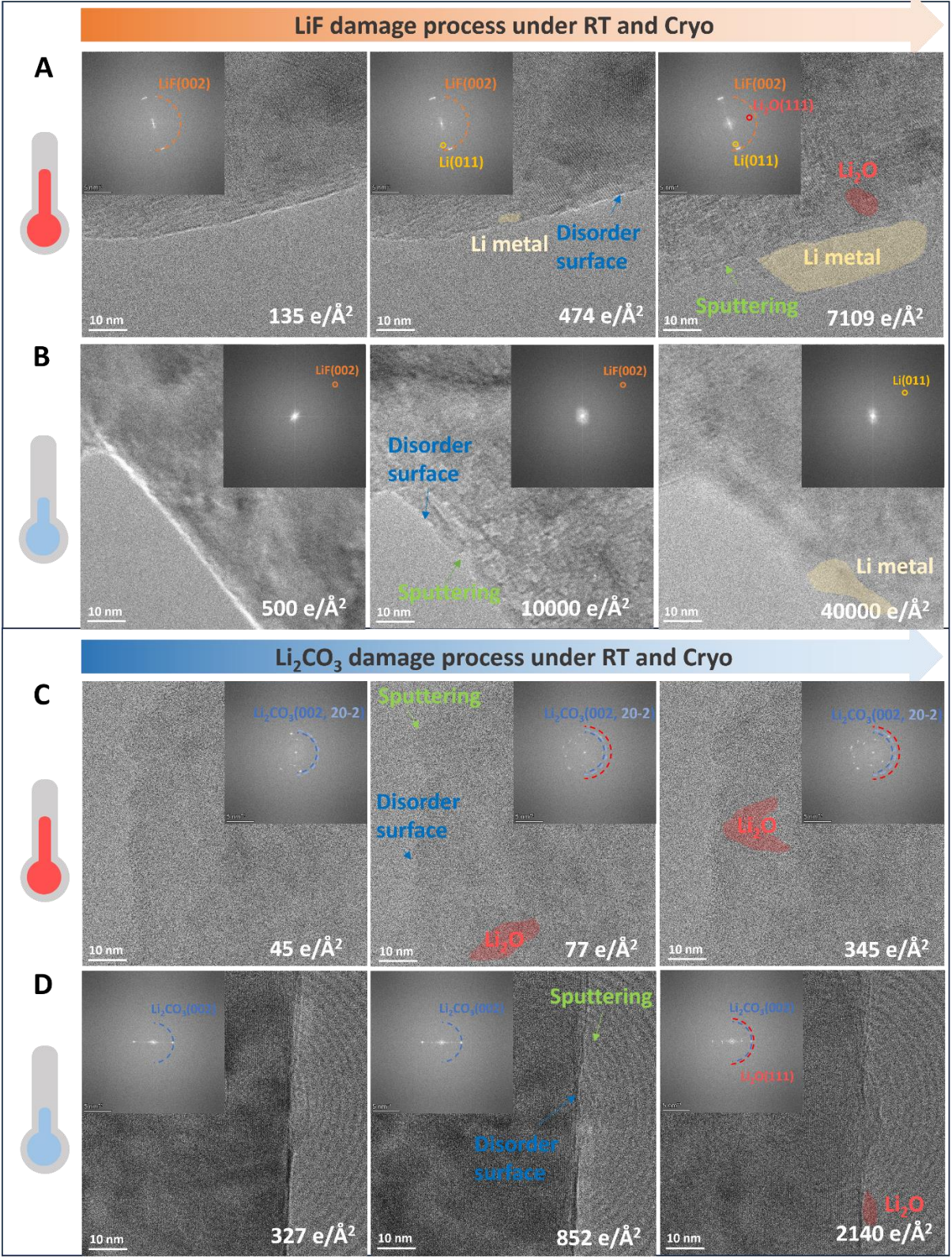

**Figure 3.** Electron Beam-induced damage process in LiF and $Li_2CO_3$. The damage to the LiF component at (A) room temperature (20°C) and (B) cryo temperature (< -170°C), and the damage to the $Li_2CO_3$ component at (C) room temperature (20°C) and (D) cryo temperature (< -170°C). All experimental measurements were conducted at 200 kV.

The mechanisms of electron irradiation damage can be broadly categorized into atomic displacement damage (electron-nucleus interactions) and radiolysis (electron-electron interactions).[28] Displacement damage occurs when high-energy electrons transfer energy and momentum to the nucleus potentially displacing atoms if enough energy is transferred. This effect is more common at higher accelerating voltages and can displace atomic nuclei to interstitial positions, degrading crystalline structures and creating disordered regions or if near a surface/interface sputter material out of the specimen. As shown in **Figure 3**, this damage to SEI components is evident in the early stages of imaging, especially at room temperature. Notably, when high energy scattering occurs in thin specimens or at their surface, the energy required for displacement is significantly reduced, leading to surface sputtering. Here we illustrate the time evolution and change of the structure of the SEI components under controlled irradiation experiments. Documented in this figure is the morphological change in high resolution TEM images (HREM) with increasing dose at RT and when cooled to $LN_2$ temperatures. The insets in each figure showing Fast Fourier Transforms taken of the HREM images and allow changes in the spacing of phase contrast lattice images to be tracked as a function of dose. This combined with traditional electron diffraction can be used to quantify changes to the phases present in the irradiated area.

Additionally, radiolysis occurs due to the interaction between incoming electrons and specimen electrons, which can cause Joule heating, bond disruption as well as chemical reactions.[29] In the case of LiF, electron beam irradiation reduces the lithium fluoride, resulting in the formation of lithium metal (**Figure 3A**). This reduction produces metallic lithium nanoparticles within the damaged area. After that, as seen in **Figure 3A**, these *in-situ* formed Li particles, due to their ultra-small size, readily react with residual oxygen in the TEM column, forming $Li_2O$. Under cryogenic conditions (**Figure 3B**), LiF still decomposes into lithium metal but with significantly higher dose tolerance. The decreased $Li_2O$ formation is observed as well. The reduction in $Li_2O$ under cryogenic conditions can be attributed to the lower initial amount of *in-situ* formed metallic lithium and its reduced oxidation activity at low temperatures. For $Li_2CO_3$ (**Figure 3C**), decomposition leads to the formation of $Li_2O$ and the presumed release of $CO_2$ as a byproduct—a process characteristic of radiolysis that occurs even under low dose rate conditions (15.9 e/Å²·s) at room temperature. Under cryogenic conditions, a similar damage process is observed, but with > 300 times higher beam tolerance (**Figure 3D**).

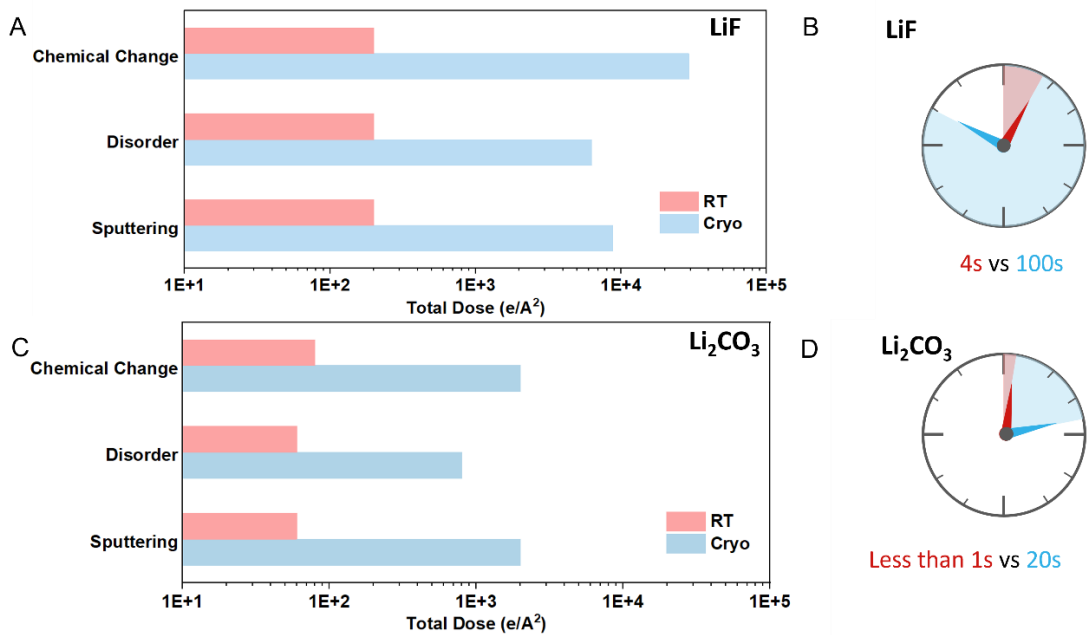

**Figure 4.** Electron beam irradiation damage mechanism and threshold of LiF and $Li_2CO_3$. (**A**) Critical total dose of damage process and (**B**) calculated operation window for LiF at room and cryogenic temperature at dose rate of 100 e/Å²·s. (**C**) Critical total dose of damage process and (**D**) operation window for $Li_2CO_3$ at room and cryogenic temperature at dose rate of 100 e/Å²·s. The bar plots in A and B represent total dosage region without noticeable damage. The time lapse indicators in B and D represent the relatively safe time intervals for imaging at room and cryogenic temperature. All datasets were acquired at 200 kV.

At a given acceleration voltage, the extent of electron beam damage is influenced by both the beam dose rate and the duration of beam exposure, which collectively determine the total beam dose. Understanding the dependence of electron beam damage on these factors is crucial for characterizing and mitigating the detrimental effects caused by irradiation. Higher beam dose rates can lead to increased damage due to the larger number of incident electrons interacting with the sample per unit time. Similarly, longer beam exposure times result in cumulative electron interactions and can escalate the overall damage inflicted upon the specimen. **Figure 4A** and **Figure 4C** thus summarize the total beam dose limitations for the electron-reactive battery materials collected in this study at both room temperature (20°C) and cryo temperature (-170°C). The threshold increases significantly from $Li_2CO_3$, LiF, to $Li_2O$ and lithium metal, which well explains that electron beam damage products in the above-mentioned standard samples are either $Li_2O$ or Li metal.

The differences in total dose thresholds arise primarily from distinct material properties, such as bond strength and atomic weight of the constituent atoms. Bond strength can be estimated by decomposition enthalpy because it reflects the energy required to break a chemical bond. When a compound decomposes, its constituent bonds are broken, and the energy absorbed during this process is represented by the decomposition enthalpy. The decomposition enthalpy for LiF is calculated to be 616.0 kJ/mol, which is much higher than that of $Li_2CO_3$ (226.7 kJ/mol) based on the following reactions:

$$LiF\ (s) = Li\ (s) + \frac{1}{2}F_2\ (g)$$

$$Li_2CO_3\ (s) = Li_2O\ (s) + CO_2(g)$$

Stronger bonds require more energy to cleave, resulting in higher decomposition enthalpies. Comparing the magnitude of decomposition enthalpies provides insights into why $Li_2CO_3$ decomposes at a significantly lower total electron dose. While changing the acceleration voltage does not notably enhance the dose limit for imaging the SEI component at room temperature (**Table S1** and **S2**), our observations in **Figure 4** highlight the critical role of cryogenic temperatures in mitigating electron damage to SEI components. Employing a specimen holder cooled by liquid nitrogen, we demonstrate a substantial enhancement in the total dose limit, improving it by two orders of magnitude.

As calculated in **Figure 4B** and **Figure 4D**, if we take the image at low dose mode of 100 e/Å$^2$·s dose rate, the operation time window for LiF is 2s at room temperature and 100s under cryo-condition, while for $Li_2CO_3$ is less than 1s and 20s, respectively. Note the damaging processes of both LiF and $Li_2CO_3$ were recorded using low dose technique (<20 e/Å$^2$·s) to mitigate any potential damage during the periods of searching and focusing for areas of interest. This low electron dose rate for atomic resolution imaging is beyond the capability of conventional CMOS camera acquisition. Essential to achieve this will be cryogenic resources configured for operation at low dose (<100 e/Å$^2$·s) using direct electron detectors.

**Standard Protocols for Correlative Imaging of Reactive Lithium Metal**

The studies above highlight the importance of meticulously controlling and reporting detailed cryo-TEM experimental conditions for lithium materials to ensure data reproducibility and comparability across different research groups. Accurately reflecting the actual battery materials/samples storage environment is crucial. Furthermore, it is essential to include comprehensive information on experimental conditions, especially the electron beam energy, as well as the dose and dose rate received by the battery sample, in the experimental section.

Since the publication of the first cryo-EM articles on lithium materials in 2017, there has been continuous growth and increasing attention toward applying cryogenic electron microscopy in energy materials research (**Figure 5A**). This technique has been utilized not only for lithium-ion batteries but also for a wide range of systems beyond lithium-ion, including sodium-ion batteries, lithium-sulfur systems, solid-state electrolytes, and polymer studies. Among these studies, we carefully identified 66 key papers that encompass nearly all significant research on cryo-EM applied to lithium metal battery materials in both liquid and solid electrolyte systems (with references listed in Supporting Information, **Table S3** and **S4**). As shown in **Figure 5B**, a substantial portion of these studies did not report the beam dose used during high-resolution imaging. Specifically, over 80% of studies on solid-state systems and nearly 60% of those on liquid electrolyte systems omitted this crucial detail. Among the studies that did report beam dosage, only 2 on solid-state systems and 11 on liquid systems employed low-dose conditions. This indicates that less than 20% of published cryo-EM studies in the context of battery research might control the beam-induced sample damage/change.

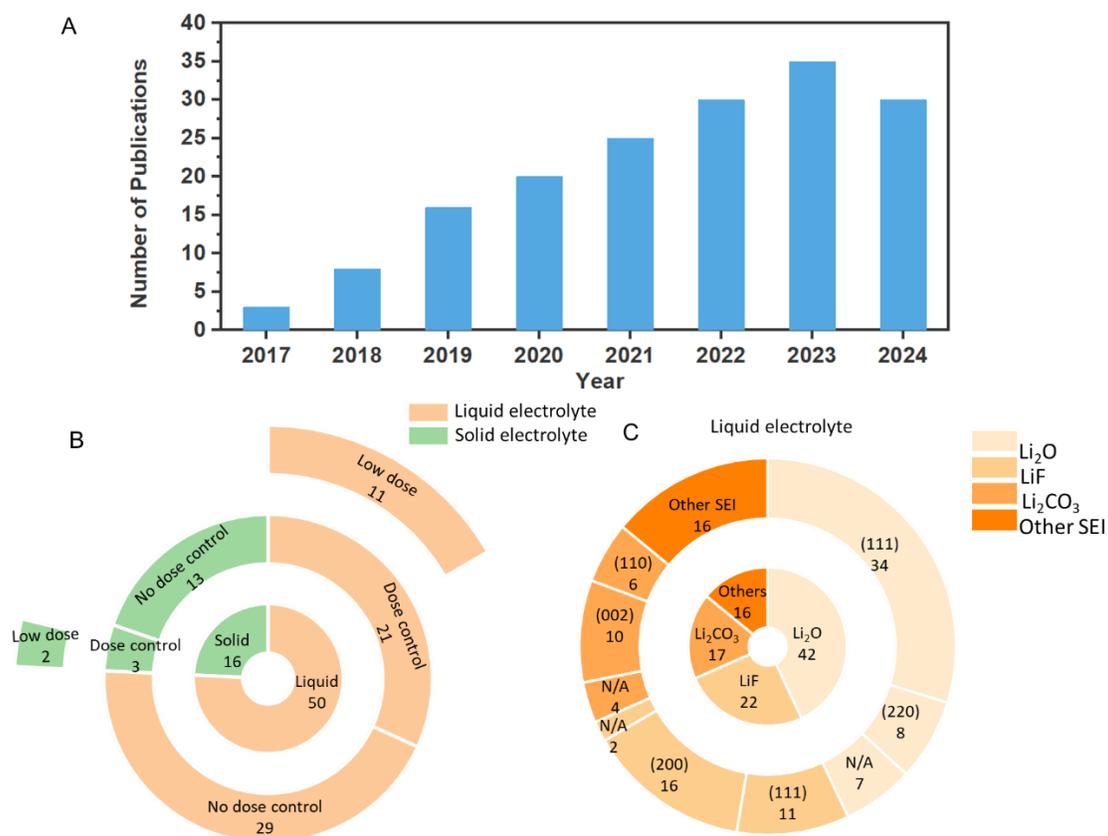

**Figure 5.** Literature summary of research associated with (S)TEM study on the SEI in battery materials. (**A**) Numbers of publications each year since the introduction of cryogenic electron microscopy into battery research field. (**B**) Numbers under each label represent the number of publications on cryo-EM analysis for lithium metal battery study. (**C**) Numbers of SEI components and crystallographic lattices identified in lithium metal batteries with conventional organic solvent-based electrolytes. Here we classify low dose to indicate TEM imaging acquisition at the electron dose rate less than 100 e/Å$^2$·s.

**Figure 5C** focuses on the liquid electrolyte systems and conducts a detailed examination of the number of reported SEI components and the crystallographic lattices reported. This figure shows $Li_2O$ is the most widely reported SEI component (>35%), followed by LiF, $Li_2CO_3$ followed by other SEI components such as LiOH, $Li_2S$, etc. The reported crystallographic phases for various SEI components align with the structural factors of different orientations and is used by the authors to indicate detection of the various crystalline forms present in those studies. Even though $Li_2O$ is mostly observed within SEI, very few publications claim the functions and properties of $Li_2O$ as the interphase material. The origin of these $Li_2O$ remains uncertain; it could be an intrinsic feature of the SEI in certain systems. However, as indicated by our data, it is also highly probable that it arises from the electron beam-induced degradation of other SEI components, such as LiF or $Li_2CO_3$. This indicates the critical importance of conducting quantitative analysis on the distribution of interphase materials using appropriate imaging, diffraction and spectroscopic methods.

The literature review highlights a gap in guidelines for the preparation, storage, transfer and characterization of beam-sensitive battery materials, particularly lithium metal, during TEM studies. This has led to the misconception that cryogenic characterization alone is sufficient to maintain sample integrity. However, with careful sample preparation and minimal storage time, the adoption of best practices in data reporting—specifically in terms of temperature and dose control—will be essential. These measures will help build a reliable database, ultimately supporting the advancement of battery technology. On a positive note, if the research focus is not on SEI components or interface analysis but rather on bulk lithium metal characterization, our data suggest that TEM images with near atomic resolution can be obtained at room temperature. This can be achieved by paying close attention to sample preparation and transfer processes.

In addition to the urging of reporting detailed experimental conditions for cryo-EM, processing the resulting high-resolution images from SEI studies can also present significant challenges due to the nature of the complexity of materials in the system and large amount of data produced. The complex materials characterization technologies embodied in the cryogenic imaging and analysis protocols is an inefficient method for comprehensive characterization of energy materials when driven by a human. Site-to-site variation and the sample preparation procedures further increase the difficulty to claim a fully comprehensive representation of material systems. To tackle this problem, a novel workflow can be applied for detecting components and phase segmentation from raw high-resolution TEM images using a deep learning trained model.[30] As shown in **Figure S6** and **S7**, the developed model can expedite the phase segmentation along the electron beam irradiation damage process for both LiF and $Li_2CO_3$, diminishing the temporal and cognitive demands associated with scrutinizing an extensive array of TEM images, thereby allowing us to apply both low dose and time-intensive analytical electron microscopy techniques.

Nevertheless, some previous studies' oversight into characterization parameter control is understandable. In the early stages of applying an emerging technology to novel materials, researchers often become excited by observations that were not possible with previous characterization techniques, sometimes overlooking the critical influence of experimental parameters on the results. Compounding this issue, it can be challenging to find alternative characterization methods to further validate these observations. However, as research progresses and cryo-EM becomes more prevalent, especially after more than seven years of application in the energy field, a thorough evaluation of processing, storage, and testing conditions is essential. This careful assessment is crucial to ensure the accuracy and reliability of future studies. Our suggested protocol ensures that subsequent studies in the field yield more accurate and reliable results. Moreover, such rigor will significantly benefit the quantitative analysis of SEI components.

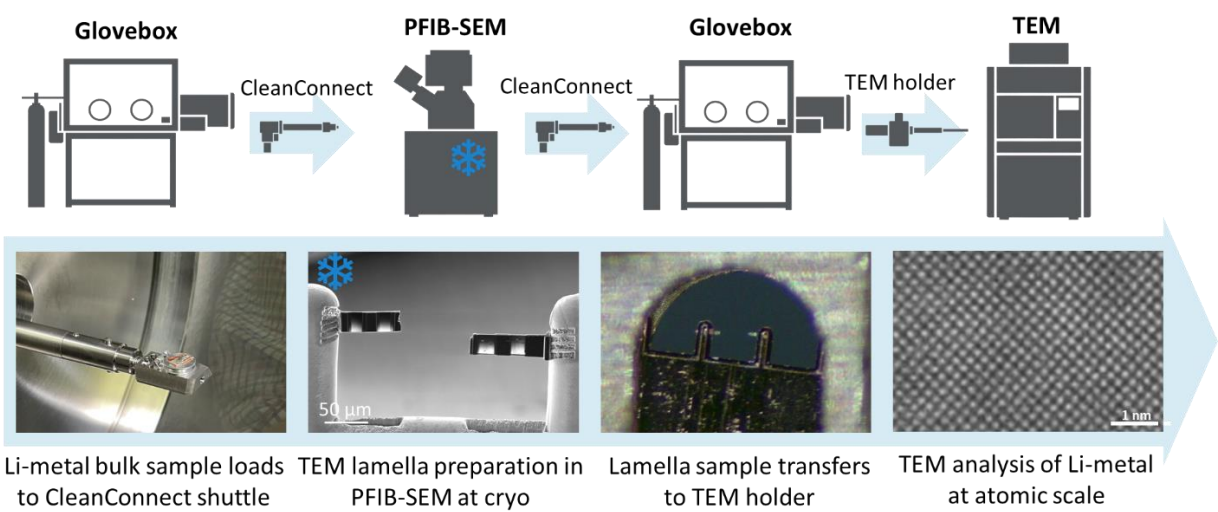

**Figure 6.** Correlative imaging workflow for PFIB-TEM characterization through inert gas sample transfer (IGST) to preserve reactive lithium metal in its native state. Snowflake icon indicates the characterization at cryogenic temperature.

Building on our understanding of lithium metal's chemical reactivity, we present a protocol for high-resolution characterization of lithium electrodes, minimizing changes from beam sources or environmental factors. This process integrates storage, preparation, transfer, and characterization for correlative imaging down to the atomic scale. As shown in **Figure 6**, the protocol starts with loading the bulk lithium sample onto the CleanConnect shuttle in the glovebox, where it is stored for up to 2 days before transfer to the PFIB-SEM. Lamella preparation, including milling, lift-out, and thinning, is done at cryogenic temperatures to prevent contamination. Multiple lamellas with two windows each are prepared for high resolution TEM, taking under an hour, compared to 5–10 hours with Ga FIB. After PFIB, lamellas are transferred back to the glovebox under Ar protection and then to the TEM via an inert gas transfer holder. Minimizing glovebox exposure is crucial, and we recommend proceeding within one day after lamella preparation to reduce contamination. The lamellas are protected under Ar or vacuum during transfer using IGST holders, ensuring high resolution imaging at room temperature, with the possibility for cryogenic SEI imaging. The imaging dose rate must be reported to verify whether beam-induced damage could affect the compositional information.

The success of the protocol relies on continuous inert gas protection, inert ion preparation, minimal glovebox storage, and the IGST holder, laying the foundation for room temperature imaging of reactive battery materials at high electron dosage rates. We understand that not all researchers may have access to the necessary equipment for such experiments, so some conditions can be moderately relaxed. For example, during sample preparation, a lower current $Ga^+$ ion beam can be used under cryogenic conditions, and a conventional cryo-transfer holder can be used for cryo-EM sample transfer. It is also beneficial to reduce humidity in the cryo-EM room to minimize ice formation that could interfere with surface imaging. However, certain aspects, such as minimizing the long-term exposure of lithium metal and prepared lamellas in the glovebox environment, and controlling the electron beam dosage, are essential for working with lithium metal-related materials.

## Conclusions

In conclusion, our research outlines a mechanistic framework for the preparation, storage, transfer, and TEM characterization of lithium metal electrode samples. For sample preparation, especially when utilizing FIB, we advocate for the use of inert gas sources such as $Xe^+$ or $Ar^+$ PFIB, as these not only enhance preparation efficiency but also mitigate the potential adverse effects of the ion source on lithium metal. Lithium metal, particularly when electrochemically prepared, exhibits high reactivity, even within a glovebox environment where trace amounts of water and oxygen can still initiate reactions. Therefore, it is imperative to minimize the storage duration of samples in the glovebox to preserve their integrity. During the transfer process, it is critical to avoid any air exposure. Our study demonstrates that by employing IGST techniques, it is possible to achieve high resolution imaging of lithium metal at room temperature without compromising the sample. For TEM characterization, although lithium metal can be imaged at room temperature, SEI components are notably sensitive to electron beam dosage and temperature.

Consequently, low-dose electron imaging under cryogenic conditions is essential to prevent the conversion of SEI components such as $Li_2CO_3$ and LiF into $Li_2O$, which could otherwise lead to inaccurate interpretations of the sample's composition and structure. To address practical concerns, we clarified ambiguities in the beam sensitivity of different SEI components and propose optimal practices for recording cryo-EM data. This protocol demonstrates broad compatibility across various experiments and is anticipated to be applicable beyond lithium metal and SEI components, to systems such as solid electrolytes and sulfur-based cathodes.

For the future, the analytical process for elucidating the dynamic beam damage mechanism necessitates innovative approaches capable of efficiently and accurately handling large datasets. Deep learning emerges as one of the potential solutions for this task. Consequently, there is a crucial need to develop coding tools equipped with deep learning capabilities to process multiple TEM images and employ suitable segmentation methods for extracting features. This enables the precise processing of crystallographic information about the specimens. Another potential solution for studying the dynamic changes includes time resolved high resolution electron spectroscopy, which is also currently under investigation. Correlative imaging and analysis, combined with advanced characterization tools such as X-ray based techniques and nuclear magnetic resonance (NMR), can offer comprehensive insights that deepen our understanding of battery materials and their behavior under operational conditions.

## Experimental Procedures

**Resource availability**

*Lead contact*

Further information and requests for resources should be directed to and will be fulfilled by the lead contact, Ying Shirley Meng (shirleymeng@uchicago.edu).

*Materials availability*

This study did not generate new, unique reagents.

*Data and code availability*

Request for the data and analysis utilized in this work will be handled by the lead contact, Ying Shirley Meng (shirleymeng@uchicago.edu).

**Materials**

$Li_2CO_3$ (99.99%) and LiF (>99.98%) were obtained from Sigma-Aldrich. These powders were carefully ground by hand using mortar and pestle for 5 minutes in Ar-filled glovebox. Li metal foil was purchased from MTI Corporation. Battery-grade lithium bis(fluorosulfonyl)imide (LiFSI) was purchased from Oakwood Products, Inc.; Bis(trifluoromethane)sulfonimide lithium 99.95% (LiTFSI) was purchased from Sigma-Aldrich. All salts were further dried at 120 °C under vacuum for 24 h before use; 1,2-dimethoxyethane (DME) anhydrous, 99.5% was purchased from Sigma-Aldrich. Solvents were dried with molecular sieves before use. LiFSI–LiTFSI were mixed in a molar ratio of 4.7:2.3 in DME to prepare the Bisalt electrolyte. Gen2 electrolyte was purchased from Gotion, Inc.

**Electrochemical Measurements**

A custom-made split cell with two titanium plungers and one polyether ether ketone (PEEK) die mold (all 1/2-inch inner diameter) was used for the Li deposition in the Li storage test. The Cu||Li cells were made by layering the Li metal foil (7 mm diameter, 50 µm thick, China Energy Lithium Co., Ltd.), Celgard 2325 separator (1/2 inch diameter) and the cleaned Cu foil between the two titanium plungers inside the PEEK die mold. Only ~5 µL of electrolyte was added to the Cu||Li cells to wet the separator. After the assembly, the split cell and the load cell were put into the cell holder, which provided the uniaxial stacking pressure at 350 kPa. The cell was tested inside the glovebox using Landt CT2001A battery cycler (Wuhan, China). Li metal was deposited to 0.25 mAh/cm$^2$ with a current density of 2 mA/cm$^2$ in both electrolytes.

**TGC**

TGC method was used to quantify the amount of inactive metallic Li formed after various storage time. After the storage, the commercial Li or deposited Li with Cu was collected and was put into a 30 mL bottle without washing. The bottle was sealed with rubber stopper to prevent gas leakage and potential safety hazards. The internal pressure of the container was then adjusted to the equilibrium of glovebox environment, whose internal pressure has been adjusted to 1 atm, with an open-ended syringe needle. After taking out the bottle from the glovebox, an excessive amount (0.5 mL) of deionized (DI) water was introduced into the bottle and H$_2$ gas was generated due to the reaction between water and reactive metallic Li in the system. The as-generated gas was then well mixed by shaking and 30 µL mixed gas was then injected into Nexis GC2030 Gas Chromatograph (Shimadzu) for H$_2$ measurement. A pre-established H$_2$ calibration curve was used to calculate the mass of metallic Li by measuring H$_2$ peak area. At least three samples were tested for each measure point by TGC.

**PFIB-SEM**

The PFIB-SEM work was conducted on a Thermo Scientific Helios 5 Hydra UX DualBeam. For the chemical reactivity analysis study using $Xe^+$ ions, the cross-section of the commercial Li metal (MTI Corp.) was prepared using 30 kV ions, starting from 1 µA for bulk cross-section followed by 200 nA and 60 nA for cleaning cross-section. A tungsten protective cap was prepared before cross-section milling, and rocking mill was applied to improve the cross-section cutting quality. The SEM imaging was collected via TLD-SE detector at 2 keV and 0.4 nA. EDS was collected at 5 keV via Oxford Ultim Max EDS detector.

For the chemical reactivity analysis using $Ar^+$ ions, the cross-section of the commercial Li metal (MTI Corp.) was prepared using 30 kV, starting from 4 µA for bulk cross-section followed by 2 µA and 120 nA for cleaning cross-section. The SEM imaging was collected via TLD-SE detector at 2 keV and 0.4 nA.

For the cross-section of commercial Li metal prepared at cryogenic temperature of -170 °C, a tungsten capping layer was deposited on the sample surface before cross-section using 30 kV $Xe^+$ ions, starting from 1 µA for bulk cross-section followed by 200 nA and 60 nA for cleaning cross-section.

**Standard Protocols for Correlative Imaging of Reactive Li Metal**

Sample loading in glovebox and transfer

Li foil stored in a glovebox was first glued onto an aluminum stub with double-sided copper tape. To ensure Li freshness, the top surface of the glued Li foil was scraped with a razor blade inside the glovebox. The mounted Li foil sample was then loaded onto a ThermoFisher Scientific CleanConnect transfer shuttle together with a small TEM grid holder, which holds a 3 mm TEM grid for FIB lamella. The CleanConnect transfer shuttle then transferred sample out of the glovebox to FIB-SEM for TEM lamella preparation.

Lamella preparation in FIB-SEM

The Li metal TEM lamella was prepared via a ThermoFisher Scientific Helios 5 Hydra FIB-SEM. First, the CleanConnect shuttle was used to transfer the bulk Li foil and the TEM grid into the FIB-SEM's vacuum sample chamber without exposure to atmosphere. To stabilize Li metal during milling, the whole shuttle was cooled down with a ThermoFisher Scientific Cryo stage inside the FIB-SEM to a temperature of < -170 °C. $Ar^+$ ion species were used during the entire TEM lamella making process.

A FIB current of 120 nA with an acceleration voltage of 30 kV was used to carry out rough milling. The initial Li metal chunk was made followed by cleaning with 40 nA FIB current under 30 kV and then lifted out from the bulk with a ThermoFisher Scientific cryo Easylift. The lift out needle was cooled to < -170 °C during the lift out process to minimize any heat transfer.

The lifted out Li metal chunk was then mounted to a copper TEM Grid under cryo condition for final thinning. The Li metal chunk was first thinned with a FIB current of 16 nA at 30 kV to about 700 nm in thickness. Then the FIB current was progressively reduced to 0.5 nA until the lamella reached about 300 nm in thickness. Final polishing was then carried out with a FIB acceleration voltage of 5 kV and a beam current of 200 pA. The thinned area of the lamella was then polished to electron transparency with a thickness around 150 nm and was then ready for TEM observation.

Multiple Li metal lamellas were made using the above method. Once lamellas were thinned to electron transparency, the cryo-stage was then warmed up progressively to room temperature. Over 40 min, the stage was warmed up from -178 to 20 °C. Then the prepared lamellas, together with the bulk Li sample and CleanConnect transfer shuttle, was then transferred from FIB-SEM's vacuum chamber to CleanConnect transfer capsule. The capsule was filled with Ar gas to protect the samples during transfer back to the glovebox.

Lamella transfer to glovebox and sample loading

Once the transfer shuttle arrived at the glovebox, the TEM grids with prepared Li metal lamellas were then transferred to a Mel-Build Atoms Defend TEM holder with a vacuum tweezer. The grid was securely mounted on the IGST TEM holder and the holder was then manually closed to seal the Ar environment inside the small chamber in the holder to protect the lamellas during the final transfer to TEM. The sealed IGST holder was then transferred out of the glovebox and then loaded to a ThermoFisher Scientific Talos F200X S/TEM for further analysis and imaging.

For the cryo-transfer workflow, the lamella sample was placed in a cryogenic grid box with a sample handling rod, then sealed in a metallized polyester bag. The bag was removed from the glovebox and immediately plunged into liquid nitrogen in a Styrofoam reservoir. The bag was cut open under liquid nitrogen and the cryogenic grid box removed and placed in a Gatan Elsa holder stand. The sample was then loaded into the Gatan Elsa cryo-holder under cryogenic conditions and transferred to the TEM under cryogenic conditions. The entire transfer from glovebox to TEM took ~25 min. A 7 min airlock time was used. The sample was imaged at cryogenic temperature.

TEM characterizations

The ThermoFisher Scientific Talos F200X S/TEM, outfitted with a Gatan continuum EELS and a Mel-Build IGST holder, serves as the primary instrument for Li metal lamellae characterization. Operated at 200 kV under low electron dose conditions, the instrument facilitates the acquisition of high-resolution images with minimal beam-induced specimen damage. The atomic-resolution TEM image was collected at room temperature using a Ceta 16M camera. The dose rate was about $3.5\times10^3$ e/Å$^2$ s. With exposure time of 500 ms, it gives a total dose of 1750 e/Å$^2$.

**TEM/STEM/AEM Studies**

The electron beam irradiation damage process of LiF and $Li_2CO_3$ were measured using the Analytical PicoProbe Electron Optical Beam Line at Argonne National Laboratory (ANL) which is the prototype of the ThermoFischer Scientific Spectra 300 Ultra X / Illiad analytical electron microscope. This sub-Angstrom resolution instrument was operated at 60, 200 and 300 kV at both room temperature and cryogenic condition as indicated in the main text for various aspects of this work. High Resolution TEM phase contrast images were collected on the ThermoFisher Falcon 4i camera under low dose condition at $1K^2$ to $4K^2$ pixel resolution with an exposure time ranging from 0.5 to 5s per frame in streaming acquisition (i.e. movie mode) and operating in both standard as well as the Fresnel Free Imaging Mode. Electron Diffraction measurements were conducted using a $1K^2$- $4K^2$ CetaII CMOS camera. Supporting X-ray and electron energy loss spectroscopy were conducted using the 4.5 sR collection angle XPAD and Ultrahigh energy resolution Illiad electron spectrometer installed on the instrument. Dose and dose rate measurements were performed using a Faraday cup calibrated beam current monitor together with accurate in-situ beam size measurements.

Electron damage and spectroscopy studies of Li, as well as LiF and $LiC_2O_3$ SEI materials were studied using an inert gas transfer cryogenic holder (IGST) from Simple Origin (Model 206) as well as a custom Be tipped cryo-transfer holder by Fishione Instruments (Mode 2550). SEI material samples were prepared by crushing and dry drop casting and as appropriate using an Argon glovebox system, from which they were in-situ sealed in the Simple Origin IGST holder and transferred to the TEM at room temperature. Once inserted into the electron microscope, experiments were either conducted at room temperature or cooled to -170º for cryogenic measurements.

**Acknowledgements**

The authors acknowledge funding support from the Energy Storage Research Alliance, an Energy Innovation Hub funded by the U.S. Department of Energy, Office of Science, Basic Energy Sciences under DE-AC02-06CH11357. Y. S. M. and N.J.Z. are grateful for the LDRD funding from Argonne National Laboratory. The Analytical Picoprobe Electron Microscope (ThermoFisher Scientific Spectra Ultra X Illiad) was developed and supported as part of a CRADA #01300710 and CRADA #A24560 among Argonne National Laboratory, University of Chicago and ThermoFisher Scientific and was supported in part by DOE Office of Science by Argonne National Laboratory under Contract No. DE-AC02-06CH11357, as well as in part by the National Science Foundation Major Research Instrumentation (MRI) Program (NSF DMR-2117896) at the University of Chicago. This work was also supported by the funding and collaboration agreement between UC San Diego and Thermo Fisher Scientific on Advanced Characterization of Energy Materials. Part of TEM images are collected at the San Diego Nanotechnology Infrastructure (SDNI), a member of the National Nanotechnology Coordinated Infrastructure, which is supported by the National Science Foundation (grant ECCS-1542148). The authors gratefully acknowledge technical support from Mel-Build and Simple Origin. The authors also thank Thermo Fisher Scientific Americas NanoPort for PFIB-SEM and TEM test for workflow validation and Ruijie Shao to prepare the Li-metal lamella sample for cryo transfer holder testing. T. S. M. acknowledges the support by the National Science Foundation Graduate Research Fellowship under Grant No. 214000.

**Author Contributions**

M. Z., Z. L., S. B., and Y. S. M. conceived the ideas and designed the experiments. A. S. conducted PFIB and EDS testing at both room temperature and cryogenic temperature for Li metal. B. L., D. C., and S. B. prepared samples for TGC analysis and analyzed the data. L. L., L. J., B. V. L., J. P. C., R. O., P. B., and A. B. acquired and analyzed the TEM data for Li metal. N. J. Z., S. B., G. R., and T. S. M. acquired and analyzed the TEM data for SEI components. M. Z. and Y. S. M. supervised the research. S. B., Z. L., S. W., and M. Z. wrote the manuscript. All authors contributed to the discussion and provided feedback on the manuscript.

## Declaration of Interests

The authors declare no conflict of interests.

## Inclusion and Diversity

We support inclusive, diverse, and equitable conduct of research.